\documentclass[journal=jacsat,manuscript=article]{achemso}

\usepackage{chemformula} 
\usepackage[T1]{fontenc} 
\usepackage{natbib}

\author{Mohit Singh}
\author{Neha Gawande}
\author{Y.S. Mayya}
\author{Rochish Thaokar}
\email{rochish@che.iitb.ac.in}
\phone{+91 (22) 2576 7241}
\fax{+91 (22) 2572 6895}
\affiliation[IIT Bombay]
{Department of Chemical Engineering, Indian  Institute of  Technology Bombay, Mumbai-400076.}

\title[Rayleigh breakup of a levitated charged droplet]
  {Effect of trap potential on the Rayleigh breakup of a levitated charged droplet}

\abbreviations{IR,NMR,UV}
\keywords{quadrupole trap, charged droplets, Rayleigh breakup}

\begin{document}

%
%
%
%
%

\begin{abstract}
    Rayleigh instability that results in the breakup of a charged droplet, levitated in a quadrupole trap, has been investigated in the literature, but only scarcely. We report here asymmetric breakup of a charged drop, levitated in a \textquotedblleft loose trap\textquotedblright, wherein, the droplet is stabilized at an off-center location in the trap. This aspect of levitation leads to an asymmetric breakup of the charged drop, predominantly in a direction opposite to that of gravity. In a first of its kind of study, we capture the successive events of the droplet deformation, breakup and relaxation of the drop after jet ejection using high speed imaging at a couple of hundred thousand frames per second. A pertinent question of the effect of the electrodynamic trap parameters such as applied voltage as well as physical parameters such as the size of the drop, gravity and conductivity on the characteristics of droplet breakup is also explored. A clear effect of the trap strength on the deformation (both symmetric and asymmetric) is observed. Moreover, the cone angle at the pole undergoing asymmetric breakup is almost independent of the applied field investigated in the experiments. All the experimental observations are compared with numerical simulations carried out using the boundary element method (BEM) in the Stokes flow limit. The BEM simulations are also extended to other experimentally achievable parameters. It is observed that the breakup is mostly field influenced, and not field induced. A plausible theory for the observations is reported, and a sensitive role of the sign of the charge on the droplet and the sign of the end cap potential, as well as the off-center location of the droplet in the trap.
\end{abstract}

\section{Introduction}
Charged droplets are often encountered in nature, for example, electrified cloud droplets as well as in artificial technologies such as combustion of fuel droplets, spray painting, crop spraying, and inkjet printing. Lord Rayleigh (1882) first derived a threshold value at which a charged drop exhibits instability \cite{rayleigh1882}. The mechanism suggested was overcoming of the force due to the surface tension of the liquid droplet by the repulsive electrostatic force. The first evidence of the disintegration of a liquid issuing from an electrified capillary in a sufficiently high external electric field was reported by \citet{zeleny17}, while the first charged droplet breakup in the strong electric field was investigated by \citet{macky31}. Further, the work of \citet{macky31} was supplemented by \citet{taylor64} using a suitable hydrodynamic theory. 

In most practical situations the dynamics of the breakup of a charged droplet is very fast. Hence for a detailed understanding of the droplet breakup characteristics, the droplets need to be suspended in space. The first systematic study of Rayleigh fission of a suspended drop was carried out by \citet{doyle64} where the droplet was levitated in a Millikan oil drop experiment \cite{millikan1935} and was observed to eject 1-10 smaller daughter droplets from a parent drop. Using a similar device, \citet{abbas67} obtained similar results but for a wider range of droplet radii (30-200 $\mu$m). In the Millikan oil drop setup, a continuous adjustment of DC suspension voltage was required against the change in mass and charge density. \citet{duft03} reported the undisputed images of Rayleigh breakup of a charged droplet levitated in an ideal Paul trap. The importance of their work was to produce the first of its kind images of systematically induced Rayleigh breakup in a quadrupolarly levitated charged drop. The time-lapsed images in their work exhibit symmetric breakup of a charged drop, where each image corresponds to an independent experiment. This was considered as a breakthrough in the field due to the correct identification of the onset of drop breakup in the experiments using a light scattering technique (see $\sim$ ref. \citet{duft02}). The signal from light scattering experiments was used to trigger a flash lamp and a CCD camera to capture the Rayleigh breakup events. Their methodology involved triggering the at a certain time delay after a threshold scattering signal, thereby allowing them to get the time lapsed images in different experiments, wherein various time delays were set. A remarkable sequence of breakup events (captured one image in one experiment) were reported. However, the success of their experiments also lay in the fact that their experiments had great accuracy and reproducibility in imparting an exact charge to exactly same sized drops, used in different experiments with almost the same evaporation rate. Although highly reproducible and accurate, their experiments were done on different droplets and can still account for experimental errors. A high speed imaging of the sequence of the droplet breakup events is therefore desirable for unequivocal demonstration of the Rayleigh instability and is presented here.

Moreover, the experimental evidence of the Rayleigh breakup of a levitated charged droplet, reported by \citet{duft03}, indicates that the droplet breakup occurs symmetrically via ejection of jets from the two poles of the droplet. In their experiments the symmetry in the droplet breakup was most likely a consequence of perfect levitation of the charged droplet exactly at the center of the quadrupole trap which was achieved by using an additional DC bias voltage, to balance the gravitational force acting on the droplet, in addition to an AC potential. However, in the typical electrospray experiments (see $\sim$ ref. \citet{gomez1994}) a droplet was observed to eject a single jet at one of the poles of a droplet showing asymmetric breakup in the presence of gravity. In view of this, very recently, \cite{singh2019subcritical} have levitated a charged droplet using AC quadrupole field without any additional DC bias voltage such that the weight of the drop gives a natural asymmetry to the system by levitating the droplet slightly away from the center of the trap. In this work\cite{singh2019subcritical}, a typical experiment consists of electrospraying a positively charged droplet (in the dripping mode) into a quadrupole trap that consists of two endcap electrodes which are shorted and separated by 20$mm$ and a ring electrode. An AC voltage of 4.5-11 $kV_{pp}$ and 100-500 $Hz$ frequency was applied between the end cap and the ring electrodes. A typically charged droplet, levitated off-center in a quadrupole trap, takes several minutes to evaporate and build the Rayleigh charge before it undergoes breakup. The events were manually recorded using a high-speed camera at a speed of around 150-200k frames per second (fps) for around 2-4 seconds. The video was played back to capture droplet center of mass oscillations, shape deformations as well as the asymmetric breakup of the droplet. Also, it was reported that a droplet suspended in an AC quadrupole field, and without a DC bias, was seen to exhibit the several phenomena in a typical high-speed video such as the droplet undergoes simultaneous center of mass motion and associated deformation of an otherwise undeformed spherical droplet. The droplet then undergoes an asymmetric breakup, predominantly in the upward direction. 

The highlight of the present work is the observance of these different stages in the entire process of the breakup of a charged drop in a single high-speed video. In this paper, we have further advanced our work reported in \citet{singh2019subcritical} by addressing the effect of fluid properties such as conductivity of the liquid droplet as well as the effect of the applied field and unbalanced gravity on the droplet breakup characteristics. Towards this, a first video graphic and quantitative evidence of the effect of the applied field, size of the droplets and conductivity of the droplet is reported and the observations are qualitatively compared with the calculation of the axisymmetric boundary integral simulations. The experiments and numerical prediction are found to be in fair agreement, the mechanism for the breakup is then elucidated.  

\section{Description of Experimental setup}

In the present work, a positively charged droplet was levitated in a modified Paul trap, the details of which are described elsewhere (see $\sim$ ref. \citet{singh2018surface}). The trap consists of two end cap electrodes and a ring electrode. The trap parameters namely $z_0$ (=10$mm$), distance between centre of the ring to the centre of the end cap electrode, and $r_0$ ($=10$mm), distance between centre of the ring to the inner periphery of the ring electrode, were significantly higher than those in the literature\cite{achtzehn05}. This allows enough space to so several operations simultaneously such as, illuminating the drop, introducing highly charged droplets generated by electrosprays and recording the drop deformation followed by Rayleigh breakup using high speed videography (by Phantom V12 camera) at 100-130k fps and a stereo zoom microscope (Nikon). 

To observe the Rayleigh breakup phenomenon, charged droplets were generated by electro-spraying ethylene glycol and ethanol solution (50\% v/v) and were levitated in an ED balance. The viscosity ($\mu_d$) of the droplet, measured using an Ostwald viscometer, was $\sim$0.006Ns/m and the surface tension ($\gamma$), measured using pendant drop (DIGIDROP, model DS) method as well as spinning drop (Dataphysics, SVT 20) method, was $\sim$30N/m. To increase the conductivity of the droplet an appropriate amount of NaCl was added to the solution. The droplet dynamics was given by the non-dimensional Mathieu equation,
\begin{equation}
x_i''(\tau)+c x_i'(\tau)-a_z x_i(\tau) \cos(\tau)+\frac{g}{\omega ^2}=0  \label{eq:r_z}
\end{equation} 
where,  $a_z$=$\frac{2 \text{Q} \Lambda_0 (\tau)}{\frac{\pi}{6}D_d^3 \rho_d  \omega ^2}$, $c$=$\frac{3 \pi  D_d \text{$\mu_a$} }{\frac{\pi}{6}D_d^3 \rho_d \omega}$, $\tau(=\omega t)$ was the non-dimensional time, $\omega=2 \pi f$, $f$ was the applied frequency.  $\Lambda_0$ was the intensity of quadrupole field and determined by fitting $\rho$ \& $z$ directional potential distribution data obtained by COMSOL Multiphysics to the equation of an ideal quadrupole potential, $\phi=\Lambda_0 (z^2-\rho^2/2)$, using multilinear regression method. $\rho_d$ was the density of the drop, $\mu_a$ was the viscosity of the air, $Q$ was the charge on the drop, $D_d$ was the droplet diameter, $x_i$=$z$,$\rho$. Since the droplet was levitated in the presence of pure AC field, without any DC field to balance the gravity, the last term in the equation \ref{eq:r_z} was added to account the gravity. The droplet shifted its position from the centre of the trap due to unbalanced gravity. Its centre mass, therefore, oscillates a new equilibrium position, $z_{shift}$. The time-averaged equilibrium position at a distance $z_{shift}$ was obtained from the simple force balance in the z-direction and given as $z_{shift}=\frac{2(1+c^2)g_z}{a_z^2\omega^2}$ where, $a_z$ $\sim$ $2a_r$, $a_r$ was the stability parameter in the $r$-direction. The value of charge on the droplet was obtained from either by matching the experimental center of mass oscillations with the numerical solution of equation \ref{eq:r_z} (see $/sim$ ref \citet{singh2019subcritical}) or by cut-off frequency method (see $\sim$ ref \citet{singh2017levitation}). In most of the experiments, the droplet was levitated at a critical value of stability parameter ($a_{z_c}=\sim0.45$ at $\sim0.01$). After levitating the droplet, the droplet breakup was recorded through high-speed imaging.

\section{Droplet breakup characteristics }
In a typical Rayleigh breakup process a levitated charged droplet is observed to undergo three consecutive events as follows: surface oscillations and deformation, breakup and relaxation back to spherical shape. The detailed drop breakup characteristics are shown in one of our recent publication (Singh et al., 2019). Few important observations are re-emphasized here for the completeness of the discussion.

(I) It is observed that a droplet undergoes several shape oscillations in sphere-prolate-sphere-oblate (SPSO) mode where the magnitude of prolate deformation increases progressively and diverges at the onset of the breakup. (II) The surface of the droplet oscillates with the driving frequency and simultaneously builds a charge density due to evaporation. The amplitude of surface oscillations depends on the applied voltage and increases progressively with increase in charge density. A very large amplitude of oscillation is observed prior to the breakup which serves as an indicator to the onset of Rayleigh breakup. (III) Although a time-varying AC quadrupole field is used to levitate the droplet, in 95\% of the cases, droplet breakup is observed in the upward direction.
(IV) At the maximum deformation, the droplet ejects a jet from the conical tip of the drop at the north-pole and the jet further disintegrates into several smaller progeny droplets. (V) The droplet is observed to eject a significant amount of charge (25-40\%) but negligible mass (\textless3\%) in the process. The charge loss in the breakup process is measured by the cut-off frequency method and by changing the applied frequency to measure the transient displacement of a levitated drop inside the trap ($\sim$ see ref. \citet{singh2017levitation}). Due to the ejection of charge, the destabilizing electric stresses reduce and the drop relaxes back to spherical shape through a series of shape oscillations.
 
Since the gravity associated with the mass of the drop is not balanced in the present experimental setup the droplet levitates slightly away from the center of the quadrupole trap in the vertically downward direction. Due to the shift from the centre of the trap, the droplet is acted upon by a uniform electric field ($E$) whose strength depends on the intensity of the applied quadrupole potential ($\Lambda$) and $z_{shift}$ from the centre of the trap, and is defined as $E$=4 $\Lambda_0$ $z_{shift}$). Thus the presence of $z_{shift}$ and thereby a uniform electric field modifies the electric stress distribution on the drop surface which leads to an asymmetric breakup of the drop. These observations were recently reported by us in a systematic study of Rayleigh breakup of a charged droplet levitated in a quadrupole trap. In this work, we addressed several additional issues which are relevant to understand the breakup physics as well as possible applications. The specific issues addressed in this work are:
\begin{enumerate}
	\item The mechanism of droplet breakup, and the evolution of droplet shape characterized by AR \& AD during the breakup. 
	\item Effect of quadrupole field strength on the deformation AR \& AD, gravitational $z_{shift}$, cone angle and jet diameter.
	\item Effect of conductivity of the droplet on the jet diameter.    
\end{enumerate} 

\section{Numerical simulations} 
Further, to validated the experimental observations and to understand the evolution of the electrical stresses responsible for the breakup, numerical simulations are carried out using axisymmetric boundary element method (BEM) for a charged viscous drop in the presence of positive or negative DC quadrupole potential. Since the droplet conductivity is high (\textgreater100 $\mu$$S/cm$) in most of the experiments and the surrounding medium (air) is assumed to be a perfect dielectric, the droplet is considered as a perfect conductor drop. Thus, to understand the mechanism of droplet breakup, simulations are carried out for a charged drop modelled as a perfect conductor. The details of the mathematical formulation and numerical implementation can be found elsewhere \cite{gawande2017}. The flow equations are solved in the Stokes flow limit while the Laplace equation is solved for the electric potential. The integral equation for the electric potential is modified by substituting applied electric potential in terms of $z_{shift}$ and is given by,
\begin{equation}
\phi(r,z)=\sqrt{Ca_\Lambda} [(z-z_{shift})^2-0.5 r^2 ] \label{simu_cal}
\end{equation}
Here, $Ca_{\Lambda}=\frac{D_{d}^{3} \epsilon}{8 \gamma}\Lambda^2$ is the electric capillary number where $\epsilon$ is the permittivity of air and $\gamma$ is the surface tension of the drop. The BEM simulations are carried out for all experimental parameters and the results are compared with the experimental observations of aspect ratio (AR) and asymmetric deformation (AD) at the onset of the breakup. Here, AR is defined as the ratio of the major axis ($L$) to the minor axis ($B$) while AD is defined as the ratio of the distance of north-pole ($L_1$) to distance of the south-pole ($L_2$) from the centroid of the drop. The values of $L$, $B$, $L_1$ and $L_2$ are obtained by tracking the boundary of the drop using the image processing tool of MATLAB software and ImageJ software. In the experiments, it is observed that the droplet breakup occurs in time which is $(1/4)^{th}$ the period of the applied AC cycle. Thus, the BEM simulations are carried out in the presence of either positive or negative DC quadrupole potential where $\Lambda_0$ is used as the intensity of the applied electric field.
\begin{figure}
	\begin{center}
		\includegraphics[width=0.6\textwidth]{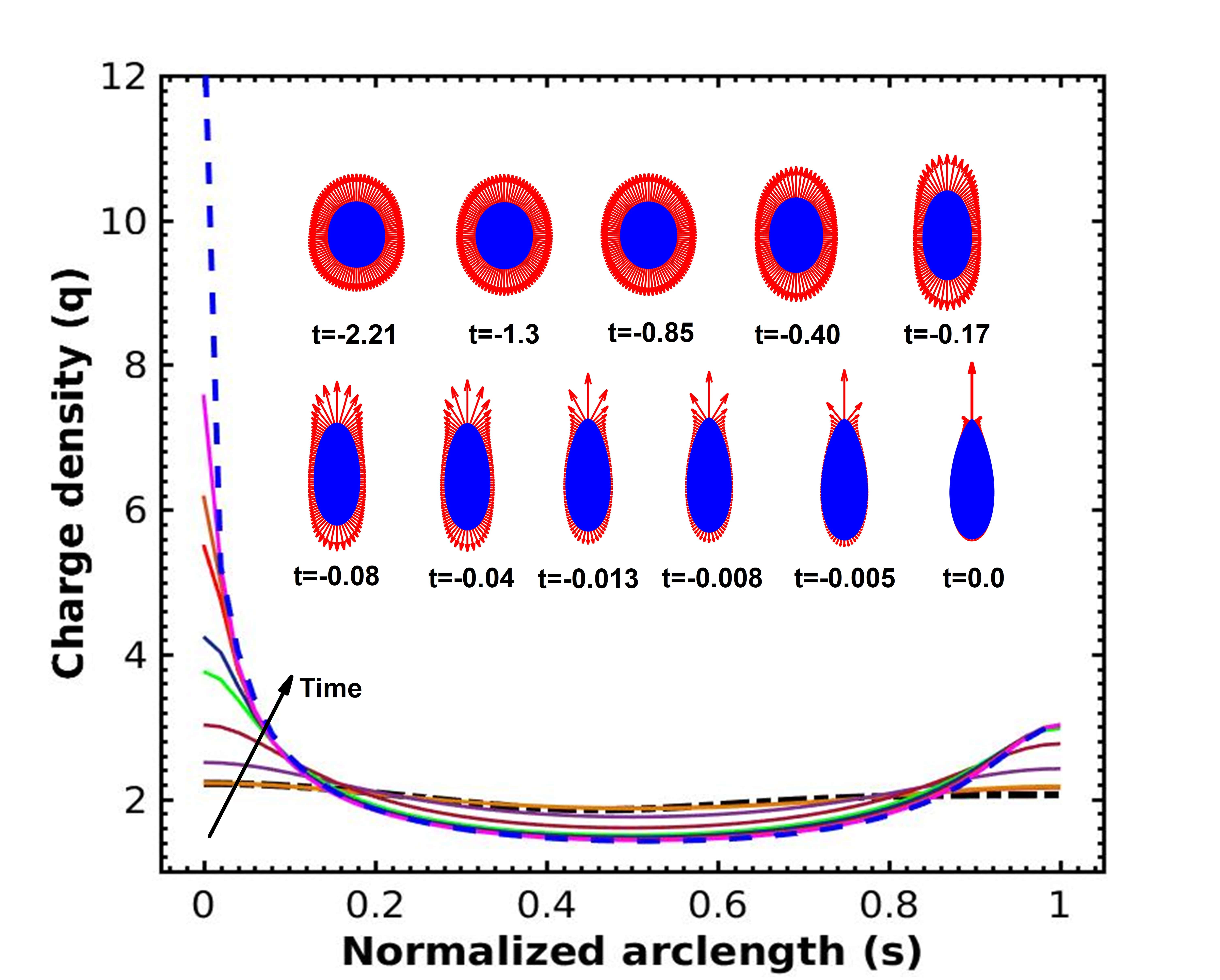}
		\caption{Surface charge density as a fucntion of arclength indicates that asymmetry is set as the drop approaches breakup. Black dash-dot line indicates the charge density distribution at t=-2.20 ms and blue dash line is at t=0 (at the onset of breakup). Inset shows the corresponding normal electric stresses obtained from the BEM simulations.}
		\label{fig:Stress_plot}
	\end{center}
\end{figure}
The simulations indicate that, in the presence of a positive DC field with $z_{shift}$, the droplet attains a stable oblate shape at equilibrium. However, in the negative DC field with $z_{shift}$, the droplet forms a conical tip at the south-pole indicating downward breakup. These results contradict the observation of upward breakup in most of the experiments. Thus the experimental results are re-analyzed and it is observed that, initially, before the instability sets in, the droplet undergoes shape oscillations. During these oscillations, the interfacial surface charge density of the droplet increases due to evaporation. As the droplet achieves a certain charge density, called as the critical charge density, the droplet deforms continuously and eventually breaks. In case of an upward breakup, the shape of the droplet at this critical point is observed to have high $P_2$ perturbation with a significant positive $P_3$ perturbation. Here $P_2$ \& $P_3$ correspond to $2^{nd}$ \& $3^{rd}$ Legendre modes. The coefficients of different Legendre modes is obtained via non-linear least-square fitting of the critical shape of the droplet using Mathematica software. Thus, in the BEM simulations, when an experimentally obtained $P_2$ perturbation is provided to the initial shape of the droplet it breaks in the upward direction when a positive DC field simultaneously acts on the droplet with $z_{shift}$. The origin of this field is the off-centered position (i.e. $z_{shift}$) due to the weight of the droplet. The droplet breaks almost symmetrically, before the positive electric field acting on $P_2$ shape and thereby charge perturbation produces a $P_3$ perturbation, inducing asymmetry and the upward breakup of the droplet. The asymmetry in the breakup is attributed to the nonlinear interaction of $P_1$ due to the uniform field and $P_2$ due to the charge on the drop which collectively gives finite $P_3$ perturbation. Thus upward or downward breakup depends on the magnitude and sign of $P_3$ perturbation. 


Further to validate the experimental observations, the shape obtained from the shape analysis of the drop at critical point, is given as an initial shape in terms of $P_2$ and $P_3$ perturbations in the simulations. The nondimensional parameters used in the simulations are as follows: $Ca_\Lambda$=0.00052, $z_{shift}$=4.5, $P_2$=0.12, $P_3$=0.02 for a droplet with $D_d$=210 $\mu$m.
The values of $P_2$ and $P_3$ are non-dimensionalized by $D_d/2$ to a the perturbed sphere of unit radius with volume $\frac{4\pi}{3}$. For the above parameters, the critical charge on the droplet required for the breakup is determined by increasing the total surface charge on the droplet with a step change of 0.1\% of the Rayleigh charge. It is found that the droplet breaks at 98.7 \% (i.e 7.9$\pi$) of the Rayleigh charge for the given parameters. This indicates that the breakup process is Coulombic and not induced by the applied field. The critical charge also indicates that the breakup is sub-Rayleigh, hinting at a subcritical instability of fine amplitude prolate perturbations. This validates the theoretical prediction of Rayleigh breakup process which shows that the charged drop exhibits transcritical bifurcation at the critical charge of 8$\pi$ \cite{das15}. 

The time evolution of the charge density and the corresponding normal electric stress acting on the surface of the drop is shown in figure \ref{fig:Stress_plot}. It can be observed that for a drop perturbed with experimentally obtained values of $P_2$ and $P_3$, the initial charge density (indicated by black dash-dot line) and the normal electric stress acting on the drop surface are slightly asymmetric where the stress is higher at the north-pole due to considerable positive $P_3$ perturbation. As time progresses the charge density at the north-pole increases rapidly as compared to that at the south pole as shown in figure \ref{fig:Stress_plot}(denoted by blue dashline). This induces asymmetric stress distribution on the drop surface with higher stresses acting at the north-pole than that at the south-pole. The charge density at the north-pole of the drop continues to increase and diverge with time as the drop approaches breakup. Thus near breakup the north-pole experiences higher electric stresses than the south-pole. To balance these high electric stresses the tip curvature at the north-pole also diverges with time and the drop breaks asymmetrically in the upward direction. This typical behaviour is attributed to the asymmetry introduced by finite amplitude of positive $P_2$ and $P_3$ perturbations which grow with time. However, it is observed that even with no initial $P_3$ perturbation, (only $P_2$ perturbation) the droplet breaks in the upward direction for given parameters. This indicates that the asymmetric Rayleigh breakup observed in the experiments is due to the up-down asymmetric redistribution of the surface charge on the droplet on account of the uniform electric field acting from the south pole to the north pole of the droplet. 

\section{Effect of various parameters}
\subsection{Effect of $Ca_\Lambda$ on deformation}

\begin{figure}
	\begin{center}
		\includegraphics[width=0.6\textwidth]{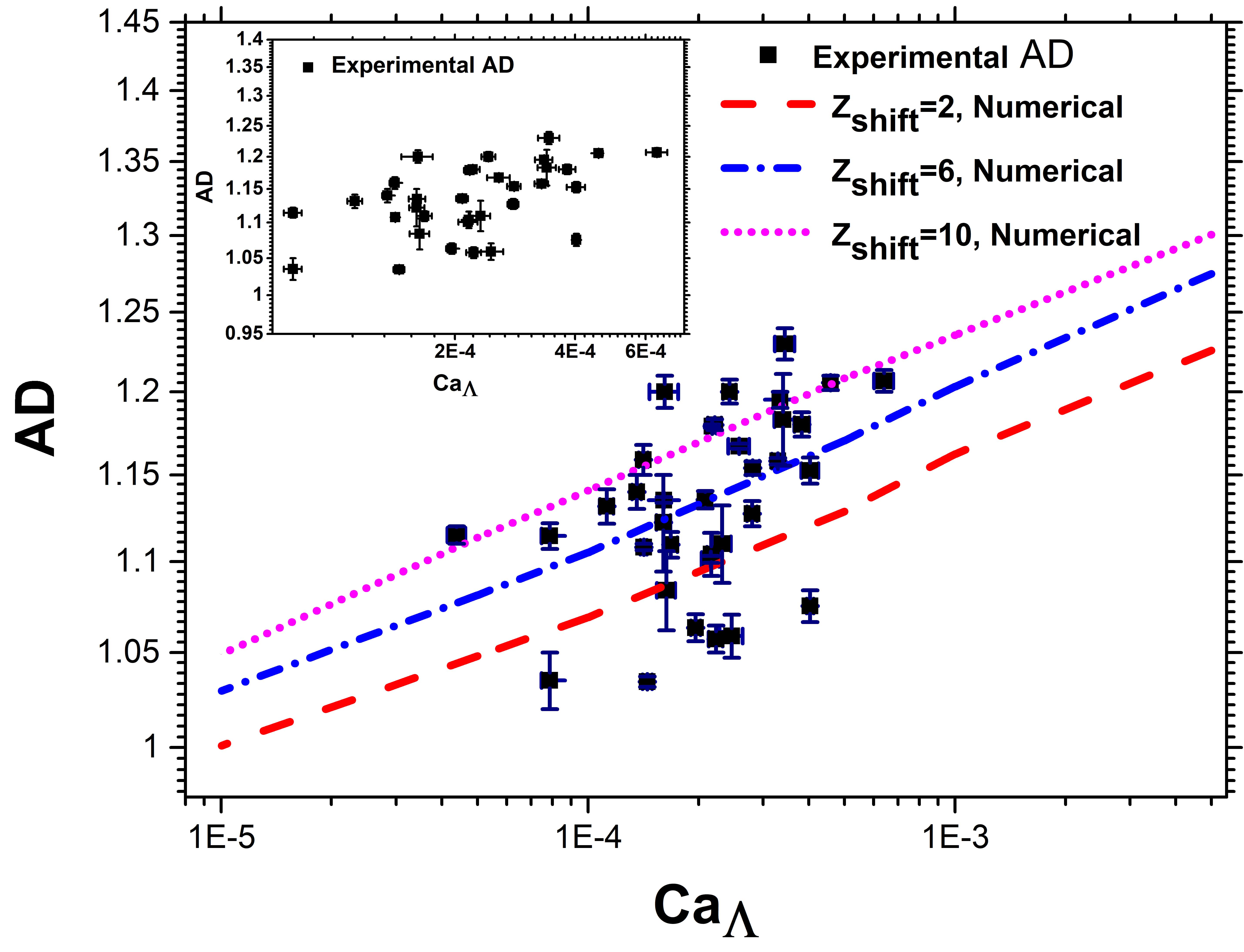}
		\caption{Effect of $Ca_{\Lambda}$ on the asymmetric deformation of the charged droplet breakup. The inset plot is the experimental observation showing the effect of $Ca_{\Lambda}$ on the AD. }
		\label{fig:AD_vs_CaL_exp_simu}
	\end{center}
\end{figure}

\begin{figure}
	\begin{center}
		\includegraphics[width=0.6\textwidth]{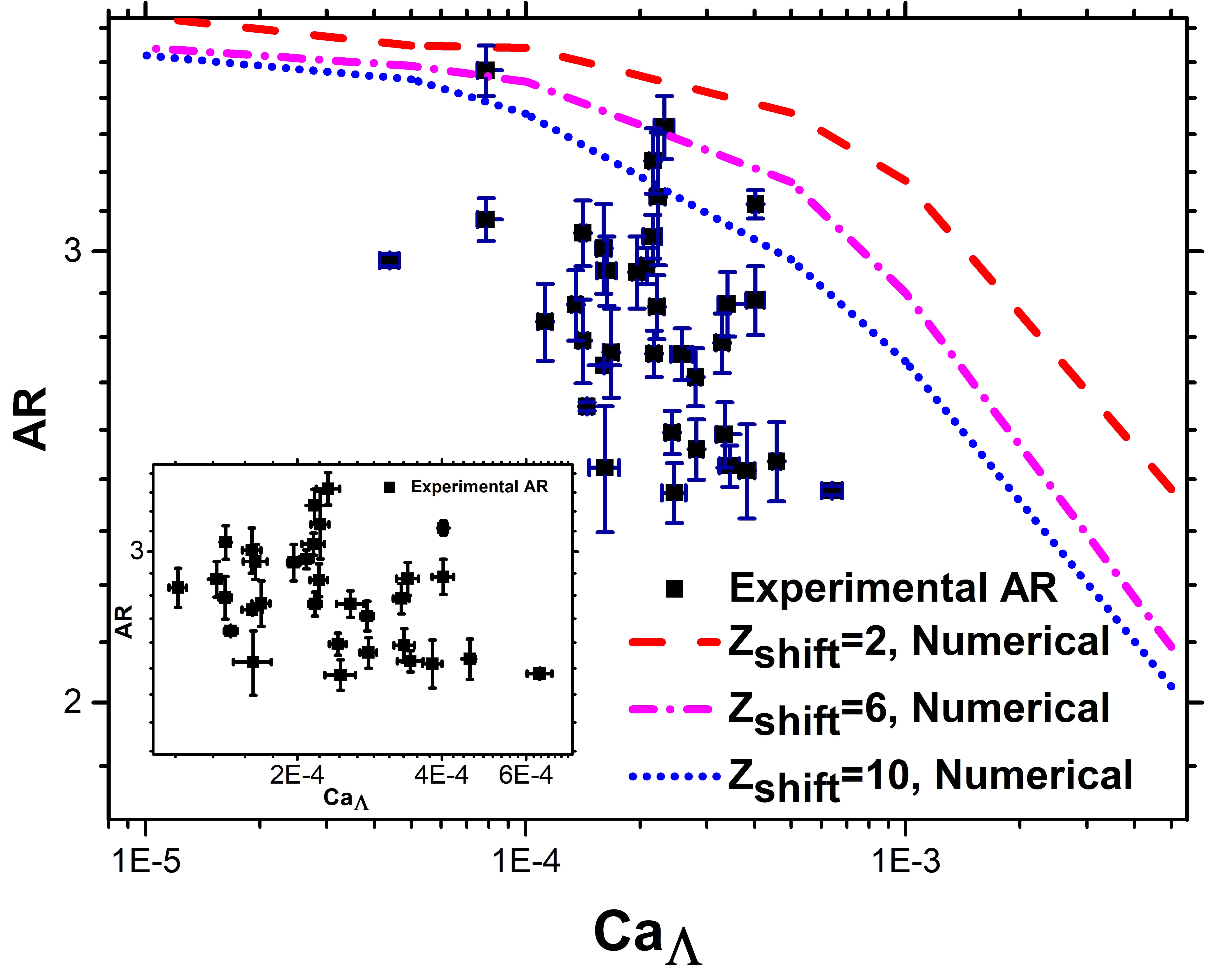}
		\caption{Effect of $Ca_{\Lambda}$ on the asymmetric deformation of the charged droplet breakup. The inset plot is the experimental observation indicating effect of $Ca_{\Lambda}$ on the AR. }
		\label{fig:AR_vs_CaL_exp_simu}
	\end{center}
\end{figure}

The magnitude of applied voltage plays an important role not only in levitating the droplet but also in the extent of the deformation prior to the breakup and thereby the subsequent droplet breakup mechanism. For example, when a bigger sized droplet is levitated at a lower voltage (4$kv_{pp}$) and higher frequency it is observed that the droplet is displaced to a greater distance from the center of the trap and exhibits very large center of mass (CM) oscillation. In this case the CM stability of the droplet is relatively poor due to lower inwardly directed time-averaged quadrupolar force ($\sim$ $\frac{(a_z^2 \bar{z}_{shift})}{2(1+c^2 )}$, where, $a_z$ is the stability parameter in the z-direction, $c$ is the drag coefficient and $\bar{z}_{shift}$ is the average downward distance from the centre of the trap (for details see \citet{ singh2018theoretical}). Since the charged droplet is levitated in the quadrupole field it is pertinent to examine the effect of applied voltage on the droplet breakup. Towards this, various values of $Ca_{\Lambda}$ are realized experimentally by changing the applied voltage used for levitating different sized charged droplets. The data discussed next is for a droplet with high electrical conductivity ($\sigma$ $\sim$ 55-1000 $\mu$S/cm). The effect of $Ca_{\Lambda}$ on the droplet breakup is characterized by measuring the values of AR and AD at the onset of the breakup, as shown in figure \ref{fig:AD_vs_CaL_exp_simu} and figure \ref{fig:AR_vs_CaL_exp_simu}. In the experiments, it is found that the values of $z_{shift}$ are different in different experiments due to different sizes of the drops at a fixed applied voltage. Moreover, the $z_{shift}$ could not be determined experimentally. Therefore, for comparison of experimental data numerical simulations are carried out for three different (non-dimensionalised by the radius of the droplet) values of $z_{shift}$=2, 6 and 10. These values are chosen in accordance with the average values observed in the experiments. The experimental values of $z_{shift}$ are estimated using the expression obtained from the $z$ directional force balance for measured parameters. 

The effect of $z_{shift}$ in BEM calculations is included by shifting the center of the trap in the positive z-direction which modifies the equation of applied quadrupole field, as shown in equation \ref{simu_cal}. The initial perturbation in the shape is given as a critical $P_2$ perturbation for a nondimensional charge fixed at Rayleigh limit (8$\pi$). The critical $P_2$ perturbation in numerical simulations is obtained by increasing its magnitude in the step of 0.01 until the droplet breaks for the given charge and $Ca_\Lambda$. In the present experimental setup $Ca_{\Lambda}$ varies from $\sim$ 4$\times$ $10^{-5}$ to $\sim$ 6$\times$ $10^{-4}$. In the numerical simulations, the values of $Ca_{\Lambda}$ are explored up to 3 decades to extend the scope of the study for other systems where $Ca_{\Lambda}$ can be of the order of $10^{-3}$.       

Figure \ref{fig:AD_vs_CaL_exp_simu} shows the effect of $Ca_\Lambda$ on the AD at the onset of breakup. In figure \ref{fig:AD_vs_CaL_exp_simu} inset, it can be observed that as the value of $Ca_\Lambda$ is increased the asymmetric deformation (AD) in the breakup increases. The error bar in $Ca_\Lambda$ is the standard deviation (SD) in the data due to uncertainty in the measurement of the droplet size. The SD of the $Ca_\Lambda$, AR and AD accounts for the error due to blurriness, camera inclination and image thresholding. The experimental results indicate that the AD values vary with $Ca_{\Lambda}$ with an exponent of 0.08. This indicates that for a given range of $Ca_{\Lambda}$, the asymmetric deformation is weakly dependent on the applied field. The simulations and experimental results (figure \ref{fig:AD_vs_CaL_exp_simu}) indicate that the BEM simulations are in qualitative agreement with the experimental observations.
\begin{figure}
	\begin{center}
		\includegraphics[width=0.6\textwidth]{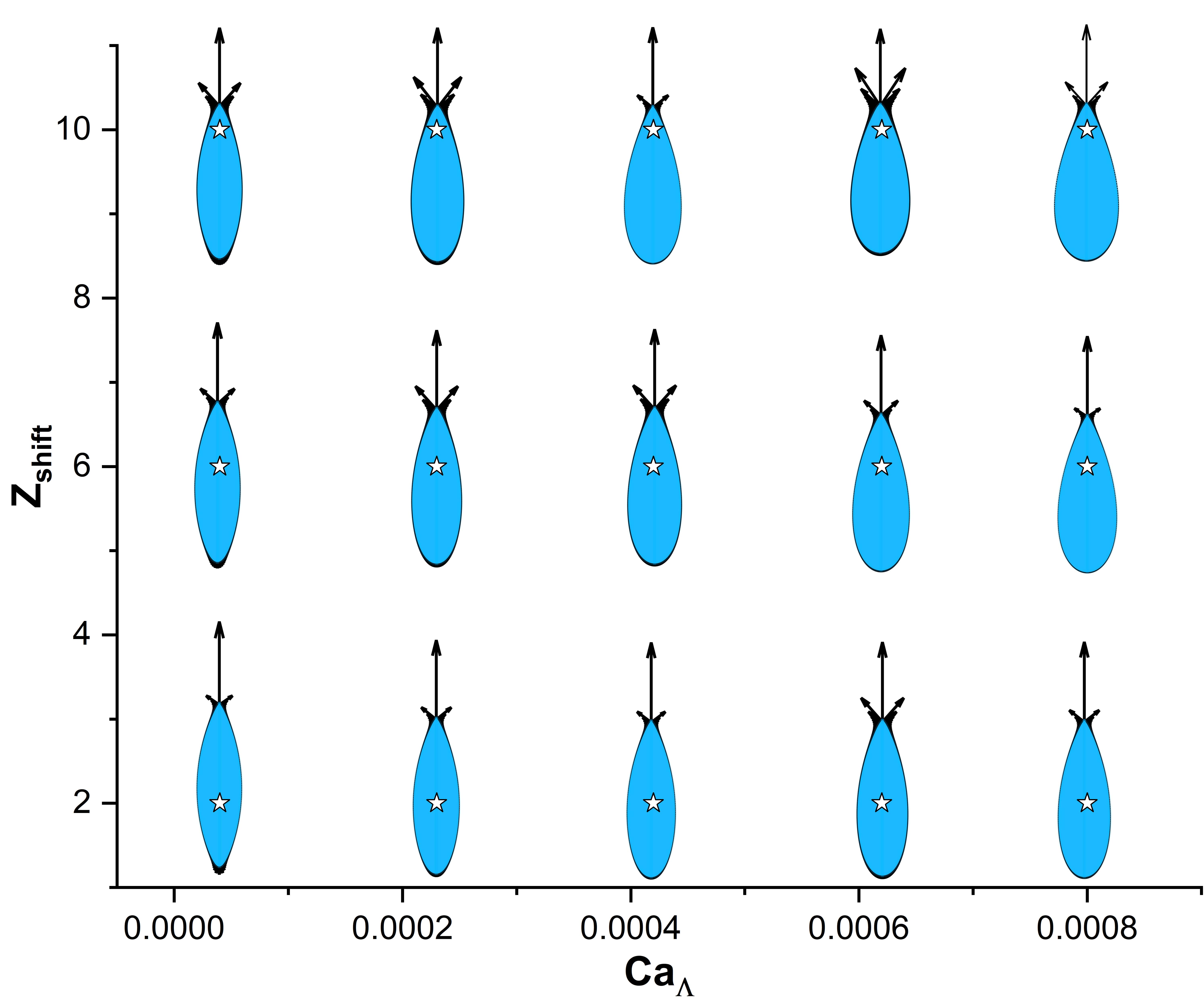}
		\caption{Normal stress distribution on the surface of the charged drop showing effect of  $Ca_{\Lambda}$ for different values of non-dimensional $z_{shift}$ i.e 2, 6, 10.}
		\label{fig:Shift_vs_caL}
	\end{center}
\end{figure}
The extent of deformation is characterized by AR and the experimental observations (figure \ref{fig:AR_vs_CaL_exp_simu} inset) show that the AR is inversely proportional, although weakly, to $Ca_{\Lambda}$. The corresponding BEM simulations support this experimental observation. Further, the electric stress distribution on the drop surface obtained from BEM simulations are shown in figure \ref{fig:Shift_vs_caL}. It can be observed that at a higher value of $Ca_{\Lambda}$, with a constant $z_{shift}$, the droplet experiences higher normal electric stresses at the north-pole than at the south-pole. This is due to the position of the droplet ($z_{shift}$) being below the center of the trap and the south endcap is considered to be at a positive potential. Thus the positively charged drop experiences high electrostatic repulsion from the positive endcap at the south-pole. This causes higher accumulation of charges in the north pole of the drop inducing higher electrical stress as compared to the south pole of the drop and introduces asymmetry in the drop deformation and breakup. When the value of $z_{shift}$ is decreased from 10 to 2 at constant $Ca_{\Lambda}$, the electrostatic repulsion from the southern end-cap is reduced and fewer charges migrate towards the north-pole of the drop. Thus the difference in the normal electric stresses acting on the two poles of the drop is reduced thereby reducing the asymmetry in the droplet breakup process. This indicates that the higher magnitude of $z_{shift}$ and $Ca_{\Lambda}$ induce more asymmetry (AD) to the shape deformation of the drop, as seen in figure \ref{fig:AD_vs_CaL_exp_simu}. The reduction in AR with the $Ca_{\Lambda}$ can be attributed to the oblate deformation tendency of the quadrupole field, thereby reducing the prolate deformation measured as AR\textgreater1. Moreover, a large value of $Ca_{\Lambda}$ leads to an early breakup, thereby reducing the AR. It should be noted that the breakup of a charged droplet can be induced by strong uniform field\cite{grimm2005dynamics}. The field influenced work as studied in this manuscript deals with the electric field of the order of 0.09 $kV/cm$ unlike the field induced breakup ($\sim$ 20 $kV/cm$) \cite{fontelos2008evolution, grimm2005dynamics}. 

\subsection{Effect of $Ca_\Lambda$ on the cone angle}
\begin{figure}
	\begin{center}
		\includegraphics[width=0.6\textwidth]{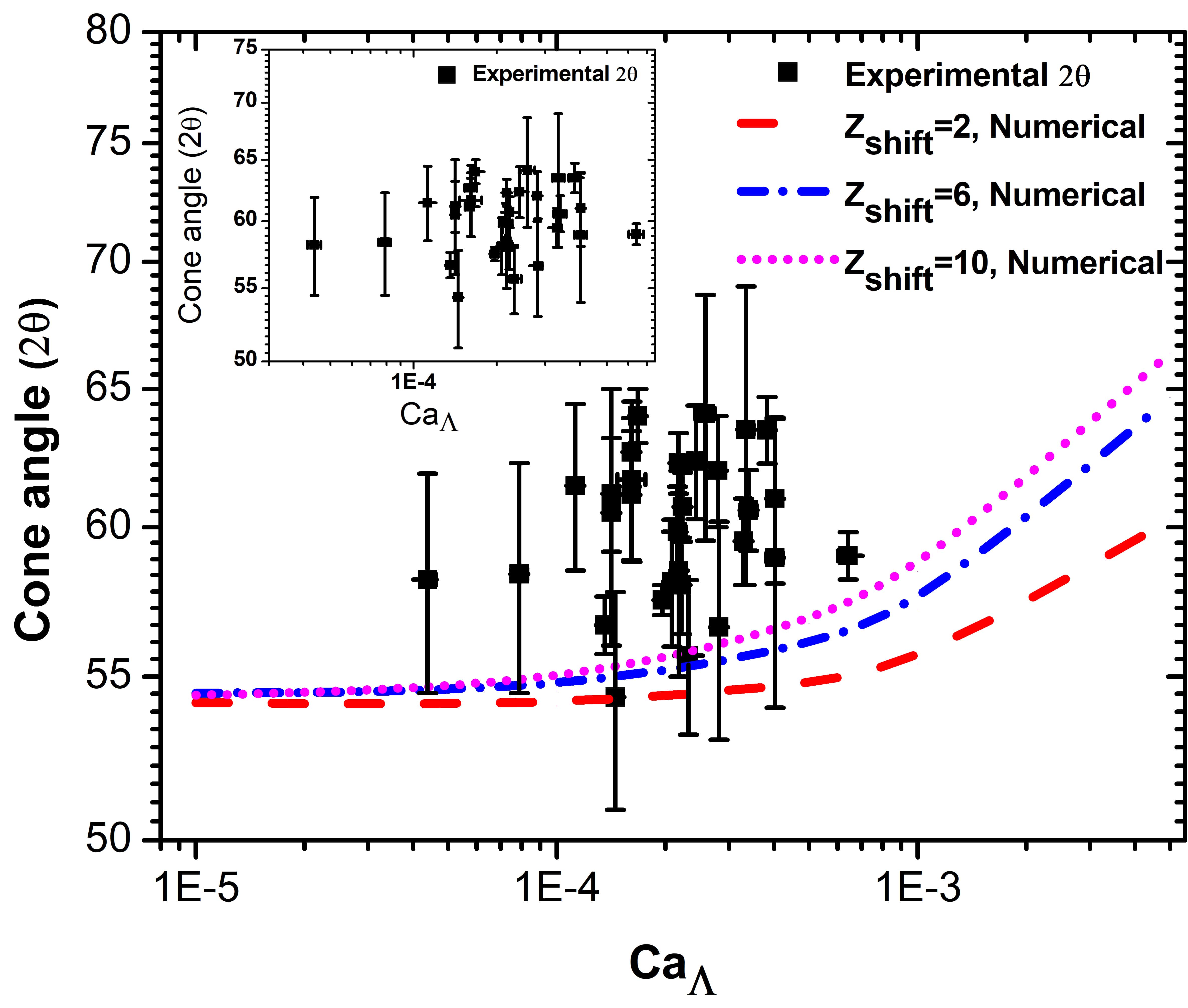}
		\caption{Change in cone angle with $Ca_{\Lambda}$. The solid vertical and horizontal lines indicate the standard deviation in the experimental data.}
		\label{fig:Cone_angle_vs_CaL}
	\end{center}
\end{figure}

The symmetric breakup of a droplet can be observed in the present setup by levitating smaller sized ($D_d$=100-170$ \mu$m) droplets at a lower voltage (4 to 7 $kv_{pp}$) thereby reducing the value of $Ca_{\Lambda}$ and $z_{shift}$. \citet{duft03} levitated a charged ethylene glycol droplet (at $80^o$C and 1 atm) at the center of a classical Paul trap and observed symmetric breakup of a charged drop. They report a semi cone angle formed during the symmetric breakup up to $30-33^o$. It is interesting to explore the effect of $Ca_{\Lambda}$ on the cone angle (2$\theta$) at a constant $z_{shift}$. Towards this the cone angle for different values of $Ca_{\Lambda}$ are plotted in the figure \ref{fig:Cone_angle_vs_CaL} and it is observed that for the experimental range of $Ca_{\Lambda}$ the value of ($\theta$) almost remains constant around $\sim$ $30^o\pm2^o$. The corresponding BEM simulations are carried out for an extended range of $Ca_{\Lambda}$ and it is observed that while the cone angle is nearly constant at low $Ca_{\Lambda}$ it increases with $Ca_{\Lambda}$ at higher values of $Ca_{\Lambda}$. This can be attributed to the field-induced effects on the breakup process. This indicates that for the $Ca_{\Lambda}$ explored in the experiments, the Rayleigh breakup process admits a nearly constant cone angle.

\subsection{Effect $Ca_{\Lambda}$ on jet diameter ($J_d$) }
\begin{figure}
	\begin{center}
		\includegraphics[width=0.6\textwidth]{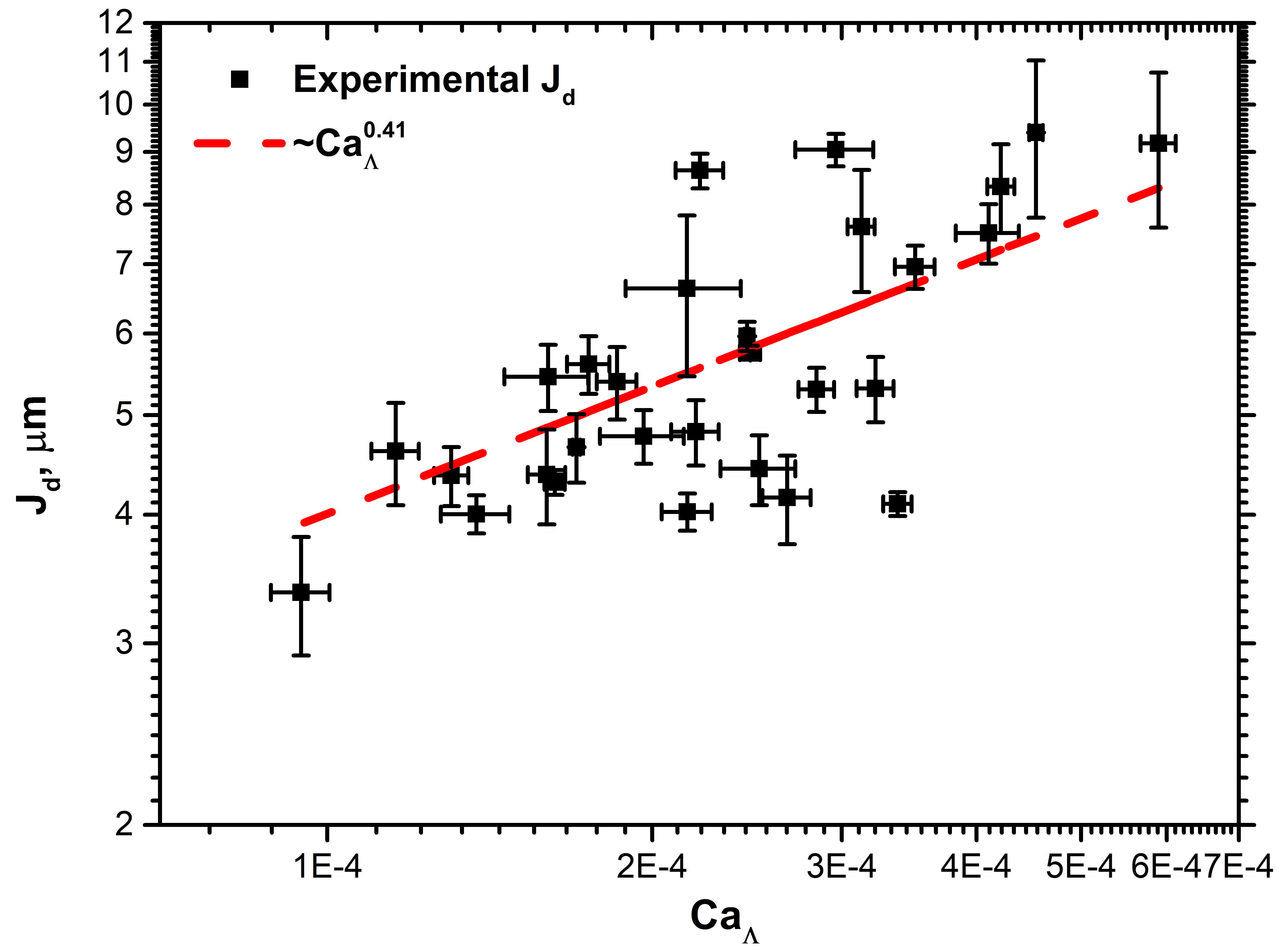}
		\caption{Effect of $Ca_\Lambda$ on the jet diameter. Parameters: $D_p$ varies from 100 to 260$\mu$m, $\Lambda_0$ varies from 2 to 3.6$\times$$10^7$V/$\text{m}^2$. }
		\label{fig:Jd_vs_caL}
	\end{center}
\end{figure}

When a droplet is levitated at high value of $Ca_{\Lambda}$ at constant $z_{shift}$ it experiences high electric stresses at the north pole and the droplet issues a jet at the north pole with higher asymmetry. It is therefore pertinent to observe how the jet diameter $J_d$ changes with the $Ca_{\Lambda}$. To examine the effect of $Ca_{\Lambda}$ a droplet of ethylene glycol and ethanol mixture is levitated. Since no NaCl is added the conductivity ($\sigma$) of the liquid drop is low (0.8-1 $\mu$S/cm). The lower conducting droplet is levitated at a fixed applied voltage i.e 11 $kV_{pp}$ and the frequency is adjusted to keep the droplet at critical stability. It is observed that for a given set of parameters the droplet ejects a thick jet at the north-pole whose thickness varies with the jet length. The $J_d$ is measured using ImageJ software and the point of measurement is chosen as the intersection of two tangents drawn at the endpoint of the cone and the starting of the jet. The different values of $Ca_{\Lambda}$ are obtained by levitating different sized drops. The change in the $J_d$ with $Ca_{\Lambda}$ is shown in figure \ref{fig:Jd_vs_caL} and it can be observed that $J_d$ increases with an increase in $Ca_{\Lambda}$ and is observed to scale as $Ca_{\Lambda}^{0.41}$. Although no direct measurement of the progeny size could be made due to poor resolution of the images the $J_d$ can be considered to be related to progeny size. Since the $J_d$ changes with the $Ca_{\Lambda}$ the sizes of the progenies are also expected to change with $Ca_{\Lambda}$. This observation contradicts the prediction of \citet{collins13}, where the simulations of a drop undergoing breakup in the presence of uniform field indicate that the deformation and subsequent progeny formation is independent of the size of the mother droplet.

\subsection{Effect of conductivity on $J_d$ }
\begin{figure}
	\begin{center}
		\includegraphics[width=0.6\textwidth]{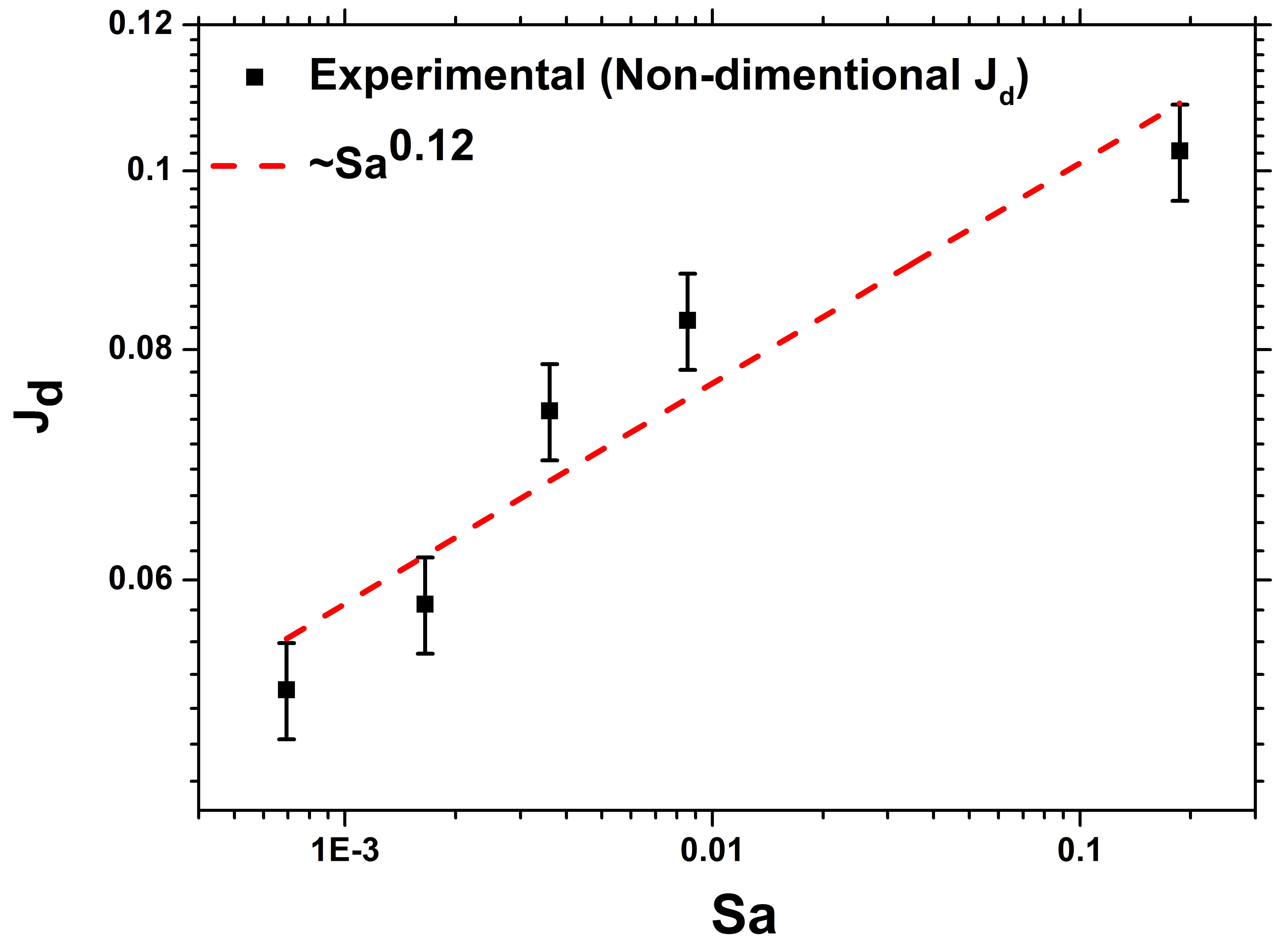}
		\caption{Change in jet diameter with Sa. Parameters: $\sigma$ varies from 1 to 250 $\mu$S/cm, $\mu_d$=0.006Pa-S, $\gamma$=0.03mN/m, $D_p$$\sim$290$\mu$m,  $\Lambda_0$=2$\times$$10^7$V/$\text{m}^2$}
		\label{fig:Jd_vs_sa}
	\end{center}
\end{figure}

A continuous jet with measurable jet thickness can be observed for low conductivity droplets. To the best of our knowledge, no experimental study is available which explores the effect of conductivity on the breakup characteristics of a levitated charged droplet. It should be noted that \citet{collins13} have looked at the effect of conductivity on the progeny size generated by the stretching of a liquid pool by a strong electric field and not due to an inherent surface charge on the droplet. Hence in the present experimental study, the effect of conductivity is explored in terms of non-dimensional Saville number (Sa) which is the ratio of charge relaxation time scale $t_e$(=$\frac{\epsilon_i}{\sigma_i}$, where $\epsilon_i$ is the permittivity of the drop) to the the hydrodynamic timescale $t_h(=\mu_i D_d/2\gamma)$ where $\mu_i$ and $\gamma$ are the viscosity and surface tension of the drop. The definition of $Sa$ suggests that the charge relaxation time is lower if the conductivity of the solution is higher. Thus the distribution of the charges over the droplet deformation time scale is instantaneous higher normal electrical stresses than the tangential electric stresses on the drop surface. In the present work, the effect of the conductivity is explored over two decades of change in the value of Sa, as shown in figure \ref{fig:Jd_vs_sa}. It can be observed that the $J_d$ varies as $Sa^{0.12}$. Thus at a lower value of $Sa$ or higher conductivity, the droplet ejects thinner jets as compared to droplets with lower conductivity.

In our experimental setup, with a moderate value of $Ca_{\Lambda}$ and $z_{shift}$, it is observed that when the conductivity ($\sigma$) of the liquid drop is increased to a very high value, the jet cannot be detected by the optical resolution of the microscope-camera assembly. At a very high value of conductivity, the drop forms a sharp tip at the north pole,  from where it ejects considerable charge but negligible mass. The ejection of charge is confirmed by the direct observation of the immediate relaxation of the drop shape after jet ejection. The reason for the formation of a sharp cone at high conductivity is the reduction in the relative magnitude of the tangential stress as compared to the normal stresses.
\begin{figure}
	\begin{center}
		\includegraphics[width=0.9\textwidth]{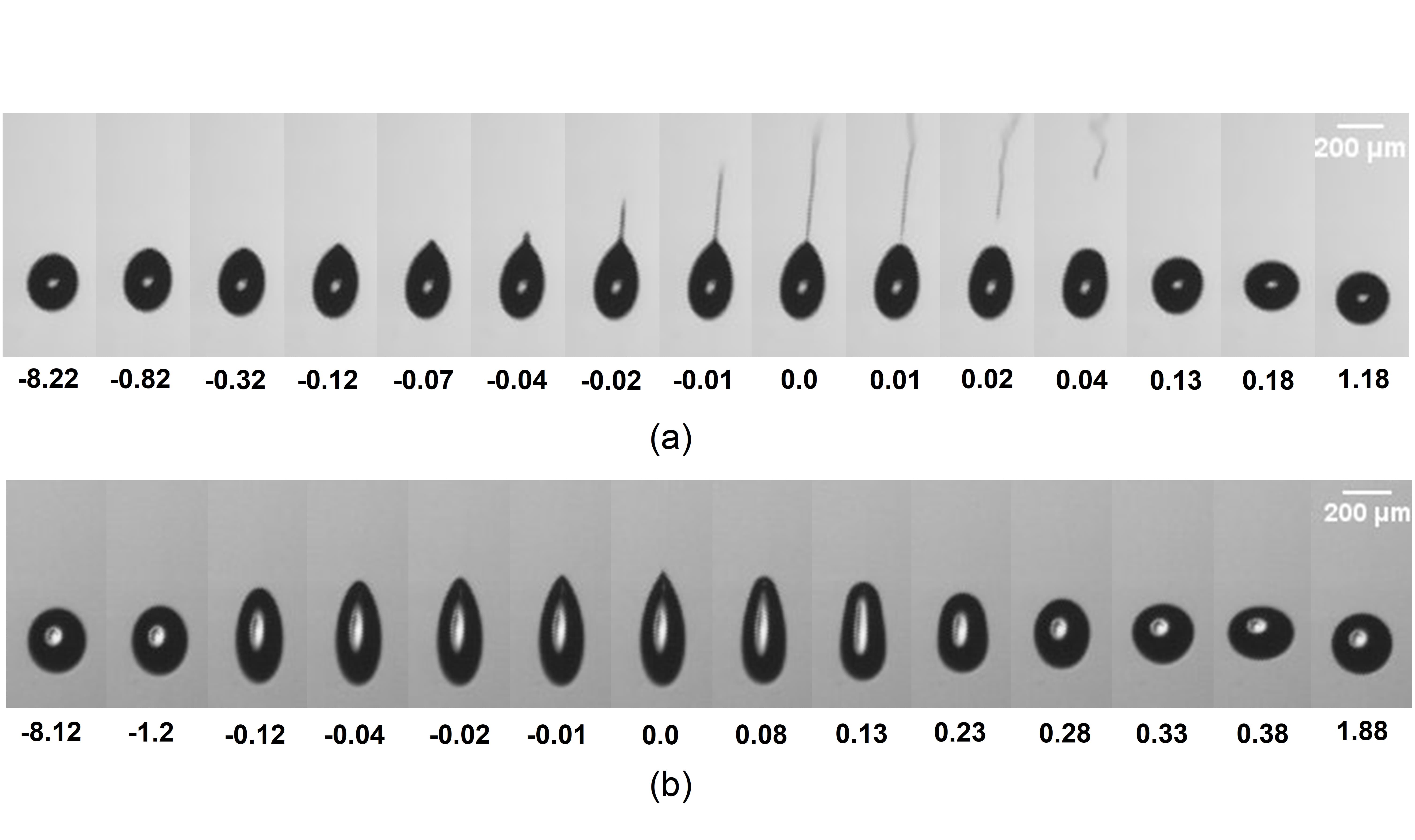}
		\caption{Deformation, breakup and relaxation sequence of droplet in the breakup process for two different conductivities. a) 1.8 $\mu$S/cm b) 1000 $\mu$S/cm. Parameters: $\Lambda_0$=2$\times$$10^7$V/$\text{m}^2$, $\mu_d$=0.006Pa-S, $\gamma$=0.03mN/m, $D_p$$\sim$200$\mu$m. }
		\label{fig:deformation}
	\end{center}
\end{figure}

\begin{figure}
	\begin{center}
		\includegraphics[width=0.6\textwidth]{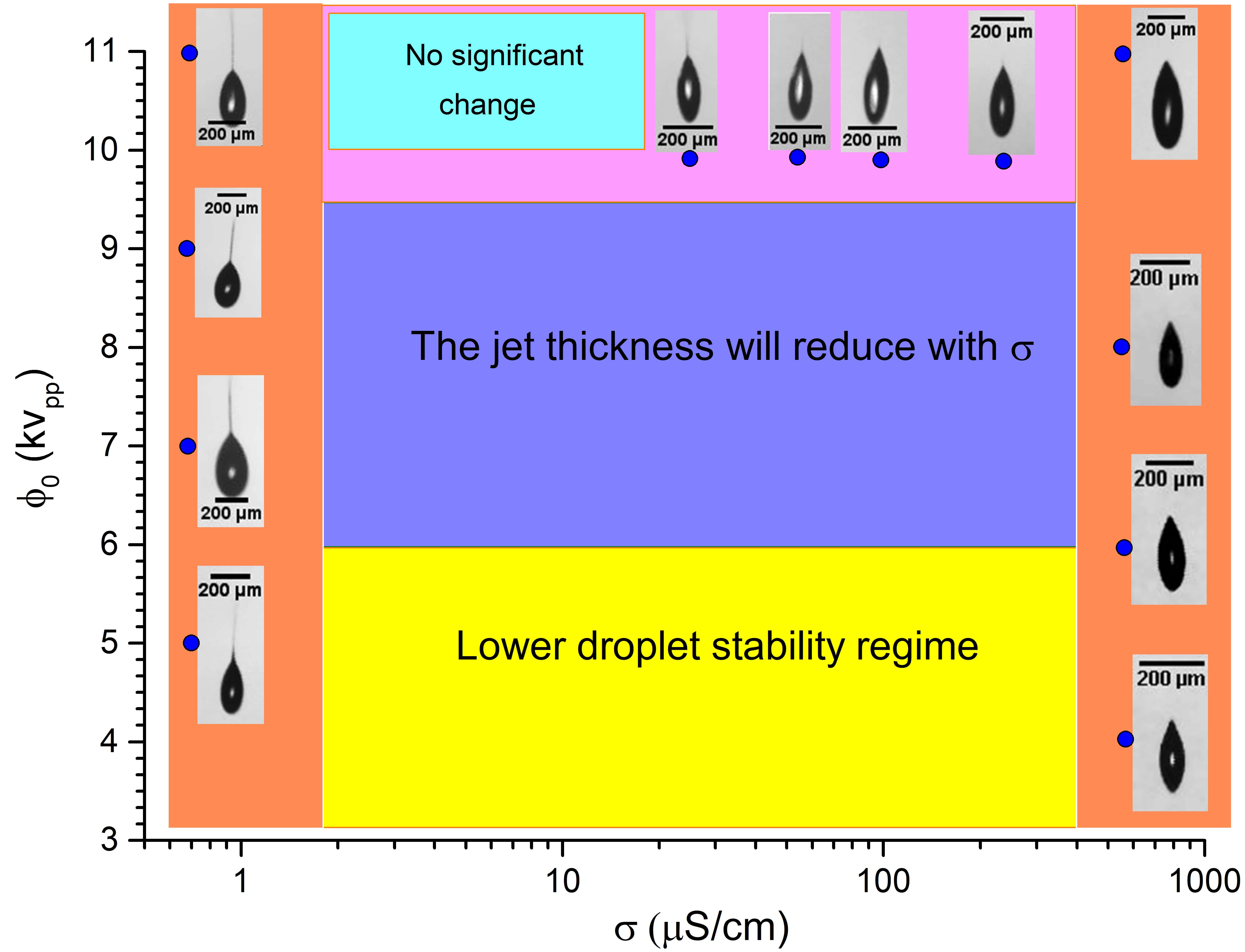}
		\caption{The phase diagram of effect of conductivity on the droplet breakup characteristics. The blue dotes corresponds to the actual coordinates in the $\sigma$-$\phi_0$ space.}
		\label{fig:phase diagram}
	\end{center}
\end{figure}

The detailed mechanism of the droplet deformation, breakup and relaxation for two conductivities (figure \ref{fig:deformation}(a), $\sigma=20\mu S/cm$, and figure \ref{fig:deformation}(b), $\sigma=1000\mu S/cm$) are shown in figure \ref{fig:deformation}. The numbers below the figures indicate the time of evolution and the unit is $\mu$s. It can be observed from the figure \ref{fig:deformation} that at time t=0, which is marked as a breakup time, figure \ref{fig:deformation}(a) shows a visible jet. On the contrary in the case of a high conductivity droplet (figure \ref{fig:deformation}(b)) a sharp conical tip is formed and the jet cannot be observed. A similar qualitative observation of jet diameter dependence on $Ca_\Lambda$ and the conductivity of the droplet is presented as a phase diagram in figure \ref{fig:phase diagram}. From figure \ref{fig:phase diagram} it can be observed that, as the conductivity of the drop increases, the jet thickness reduces and at $\sigma$ i.e $\sim$ 1000 $\mu$s/cm the droplet breaks with the formation of a sharp tip and no jet is visible. On the other hand, when the quadrupole potential is increased, at lower conductivity ($\sigma$ $\sim$ 20 $\mu$s/cm) the extent of asymmetry increases with an increase in $\phi_0$, in agreement with BEM simulations (figure \ref{fig:Shift_vs_caL}). At a higher conductivity, an increase in $\phi_0$ only leads to more oblate (and thereby fatter droplets) contribution to the droplet shape, at the breakup.    
\section{Conclusions}
The effect of applied voltage on the droplet breakup characteristics is reported for non-zero gravity and it is observed that while the AD values increase with an increase in the value of $Ca_\Lambda$   the  AR decreases. The comparison of average cone angle for the symmetric and asymmetric breakup with experimentally obtained values shows that the cone angle remains constant at about $30^0$ at low $Ca_\Lambda$, while it increases with $Ca_{\Lambda}$ above a certain $Ca_\Lambda$. Thus at moderate values of trap parameter, it can be conjectured that the breakup is indeed Rayleigh breakup, which is only influenced by the external field. On the other hand, at higher values of $Ca_{\Lambda}$, the instability could be induced by the external field. In the experiments though, the instability seems to be in the former category.  The change of the $J_d$ with the $Ca_\Lambda$ and $\sigma$ of the droplet is reported here for the first time. The magnitude of $J_d$ is observed to be higher in case of low conductivity and high $Ca_\Lambda$ which results in the formation of larger progeny droplets. The experimental observations are validated with the perfect conductor model using BEM simulations. These results further the understanding of Rayleigh breakup of charged drops in quadrupole traps, and indicates that in actual experimental and technological setups, the progeny sizes, as well as charge ejection, could be significantly affected by the electrostatic conditions in these setups.

\providecommand{\latin}[1]{#1}
\providecommand*\mcitethebibliography{\thebibliography}
\csname @ifundefined\endcsname{endmcitethebibliography}
{\let\endmcitethebibliography\endthebibliography}{}


\begin{mcitethebibliography}{21}
	\providecommand*\natexlab[1]{#1}
	\providecommand*\mciteSetBstSublistMode[1]{}
	\providecommand*\mciteSetBstMaxWidthForm[2]{}
	\providecommand*\mciteBstWouldAddEndPuncttrue
	{\def\EndOfBibitem{\unskip.}}
	\providecommand*\mciteBstWouldAddEndPunctfalse
	{\let\EndOfBibitem\relax}
	\providecommand*\mciteSetBstMidEndSepPunct[3]{}
	\providecommand*\mciteSetBstSublistLabelBeginEnd[3]{}
	\providecommand*\EndOfBibitem{}
	\mciteSetBstSublistMode{f}
	\mciteSetBstMaxWidthForm{subitem}{(\alph{mcitesubitemcount})}
	\mciteSetBstSublistLabelBeginEnd
	{\mcitemaxwidthsubitemform\space}
	{\relax}
	{\relax}
	
	\bibitem[Rayleigh(1882)]{rayleigh1882}
	Rayleigh,~L. \emph{The London, Edinburgh, and Dublin Philosophical Magazine and
		Journal of Science} \textbf{1882}, \emph{14}, 184--186\relax
	\mciteBstWouldAddEndPuncttrue
	\mciteSetBstMidEndSepPunct{\mcitedefaultmidpunct}
	{\mcitedefaultendpunct}{\mcitedefaultseppunct}\relax
	\EndOfBibitem
	\bibitem[Zeleny(1917)]{zeleny17}
	Zeleny,~J. \emph{Physical Review} \textbf{1917}, \emph{10}, 1\relax
	\mciteBstWouldAddEndPuncttrue
	\mciteSetBstMidEndSepPunct{\mcitedefaultmidpunct}
	{\mcitedefaultendpunct}{\mcitedefaultseppunct}\relax
	\EndOfBibitem
	\bibitem[Macky(1931)]{macky31}
	Macky,~W. \emph{Proceedings of the Royal Society of London. Series A,
		Mathematical and Physical Character} \textbf{1931}, 565--587\relax
	\mciteBstWouldAddEndPuncttrue
	\mciteSetBstMidEndSepPunct{\mcitedefaultmidpunct}
	{\mcitedefaultendpunct}{\mcitedefaultseppunct}\relax
	\EndOfBibitem
	\bibitem[Taylor(1964)]{taylor64}
	Taylor,~G. Disintegration of water drops in an electric field. Proceedings of
	the Royal Society of London A: Mathematical, Physical and Engineering
	Sciences. 1964; pp 383--397\relax
	\mciteBstWouldAddEndPuncttrue
	\mciteSetBstMidEndSepPunct{\mcitedefaultmidpunct}
	{\mcitedefaultendpunct}{\mcitedefaultseppunct}\relax
	\EndOfBibitem
	\bibitem[Doyle \latin{et~al.}(1964)Doyle, Moffett, and Vonnegut]{doyle64}
	Doyle,~A.; Moffett,~D.~R.; Vonnegut,~B. \emph{Journal of Colloid Science}
	\textbf{1964}, \emph{19}, 136--143\relax
	\mciteBstWouldAddEndPuncttrue
	\mciteSetBstMidEndSepPunct{\mcitedefaultmidpunct}
	{\mcitedefaultendpunct}{\mcitedefaultseppunct}\relax
	\EndOfBibitem
	\bibitem[Millikan(1935)]{millikan1935}
	Millikan,~R.~A. \emph{Electrons, Protons, Photons, Neutrons, and Cosmic Rays.};
	1935\relax
	\mciteBstWouldAddEndPuncttrue
	\mciteSetBstMidEndSepPunct{\mcitedefaultmidpunct}
	{\mcitedefaultendpunct}{\mcitedefaultseppunct}\relax
	\EndOfBibitem
	\bibitem[Abbas and Latham(1967)Abbas, and Latham]{abbas67}
	Abbas,~M.; Latham,~J. \emph{Journal of Fluid Mechanics} \textbf{1967},
	\emph{30}, 663--670\relax
	\mciteBstWouldAddEndPuncttrue
	\mciteSetBstMidEndSepPunct{\mcitedefaultmidpunct}
	{\mcitedefaultendpunct}{\mcitedefaultseppunct}\relax
	\EndOfBibitem
	\bibitem[Duft \latin{et~al.}(2003)Duft, Achtzehn, M{\"u}ller, Huber, and
	Leisner]{duft03}
	Duft,~D.; Achtzehn,~T.; M{\"u}ller,~R.; Huber,~B.~A.; Leisner,~T. \emph{Nature}
	\textbf{2003}, \emph{421}, 128--128\relax
	\mciteBstWouldAddEndPuncttrue
	\mciteSetBstMidEndSepPunct{\mcitedefaultmidpunct}
	{\mcitedefaultendpunct}{\mcitedefaultseppunct}\relax
	\EndOfBibitem
	\bibitem[Duft \latin{et~al.}(2002)Duft, Lebius, Huber, Guet, and
	Leisner]{duft02}
	Duft,~D.; Lebius,~H.; Huber,~B.~A.; Guet,~C.; Leisner,~T. \emph{Physical Review
		Letters} \textbf{2002}, \emph{89}, 084503--084507\relax
	\mciteBstWouldAddEndPuncttrue
	\mciteSetBstMidEndSepPunct{\mcitedefaultmidpunct}
	{\mcitedefaultendpunct}{\mcitedefaultseppunct}\relax
	\EndOfBibitem
	\bibitem[Gomez and Tang(1994)Gomez, and Tang]{gomez1994}
	Gomez,~A.; Tang,~K. \emph{Physics of Fluids} \textbf{1994}, \emph{6},
	404--414\relax
	\mciteBstWouldAddEndPuncttrue
	\mciteSetBstMidEndSepPunct{\mcitedefaultmidpunct}
	{\mcitedefaultendpunct}{\mcitedefaultseppunct}\relax
	\EndOfBibitem
	\bibitem[Singh \latin{et~al.}(2019)Singh, Gawande, Mayya, and
	Thaokar]{singh2019subcritical}
	Singh,~M.; Gawande,~N.; Mayya,~Y.; Thaokar,~R. \emph{arXiv preprint
		arXiv:1907.02294} \textbf{2019}, \relax
	\mciteBstWouldAddEndPunctfalse
	\mciteSetBstMidEndSepPunct{\mcitedefaultmidpunct}
	{}{\mcitedefaultseppunct}\relax
	\EndOfBibitem
	\bibitem[Singh \latin{et~al.}(2018)Singh, Gawande, Mayya, and
	Thaokar]{singh2018surface}
	Singh,~M.; Gawande,~N.; Mayya,~Y.; Thaokar,~R. \emph{Physics of Fluids}
	\textbf{2018}, \emph{30}, 122105\relax
	\mciteBstWouldAddEndPuncttrue
	\mciteSetBstMidEndSepPunct{\mcitedefaultmidpunct}
	{\mcitedefaultendpunct}{\mcitedefaultseppunct}\relax
	\EndOfBibitem
	\bibitem[Achtzehn \latin{et~al.}(2005)Achtzehn, M{\"u}ller, Duft, and
	Leisner]{achtzehn05}
	Achtzehn,~T.; M{\"u}ller,~R.; Duft,~D.; Leisner,~T. \emph{The European Physical
		Journal D-Atomic, Molecular, Optical and Plasma Physics} \textbf{2005},
	\emph{34}, 311--313\relax
	\mciteBstWouldAddEndPuncttrue
	\mciteSetBstMidEndSepPunct{\mcitedefaultmidpunct}
	{\mcitedefaultendpunct}{\mcitedefaultseppunct}\relax
	\EndOfBibitem
	\bibitem[Singh \latin{et~al.}(2017)Singh, Mayya, Gaware, and
	Thaokar]{singh2017levitation}
	Singh,~M.; Mayya,~Y.; Gaware,~J.; Thaokar,~R.~M. \emph{Journal of Applied
		Physics} \textbf{2017}, \emph{121}, 054503\relax
	\mciteBstWouldAddEndPuncttrue
	\mciteSetBstMidEndSepPunct{\mcitedefaultmidpunct}
	{\mcitedefaultendpunct}{\mcitedefaultseppunct}\relax
	\EndOfBibitem
	\bibitem[Gawande \latin{et~al.}(2017)Gawande, Mayya, and Thaokar]{gawande2017}
	Gawande,~N.; Mayya,~Y.; Thaokar,~R. \emph{Physical Review Fluids}
	\textbf{2017}, \emph{2}, 113603\relax
	\mciteBstWouldAddEndPuncttrue
	\mciteSetBstMidEndSepPunct{\mcitedefaultmidpunct}
	{\mcitedefaultendpunct}{\mcitedefaultseppunct}\relax
	\EndOfBibitem
	\bibitem[Das \latin{et~al.}(2015)Das, Mayya, and Thaokar]{das15}
	Das,~S.; Mayya,~Y.; Thaokar,~R. \emph{EPL (Europhysics Letters)} \textbf{2015},
	\emph{111}, 24006\relax
	\mciteBstWouldAddEndPuncttrue
	\mciteSetBstMidEndSepPunct{\mcitedefaultmidpunct}
	{\mcitedefaultendpunct}{\mcitedefaultseppunct}\relax
	\EndOfBibitem
	\bibitem[Singh \latin{et~al.}(2018)Singh, Thaokar, Khan, and
	Mayya]{singh2018theoretical}
	Singh,~M.; Thaokar,~R.; Khan,~A.; Mayya,~Y. \emph{Physical Review E}
	\textbf{2018}, \emph{98}, 032202\relax
	\mciteBstWouldAddEndPuncttrue
	\mciteSetBstMidEndSepPunct{\mcitedefaultmidpunct}
	{\mcitedefaultendpunct}{\mcitedefaultseppunct}\relax
	\EndOfBibitem
	\bibitem[Grimm and Beauchamp(2005)Grimm, and Beauchamp]{grimm2005dynamics}
	Grimm,~R.~L.; Beauchamp,~J.~L. \emph{The Journal of Physical Chemistry B}
	\textbf{2005}, \emph{109}, 8244--8250\relax
	\mciteBstWouldAddEndPuncttrue
	\mciteSetBstMidEndSepPunct{\mcitedefaultmidpunct}
	{\mcitedefaultendpunct}{\mcitedefaultseppunct}\relax
	\EndOfBibitem
	\bibitem[Fontelos \latin{et~al.}(2008)Fontelos, Kindel{\'a}n, and
	Vantzos]{fontelos2008evolution}
	Fontelos,~M.~A.; Kindel{\'a}n,~U.; Vantzos,~O. \emph{Physics of Fluids}
	\textbf{2008}, \emph{20}, 092110\relax
	\mciteBstWouldAddEndPuncttrue
	\mciteSetBstMidEndSepPunct{\mcitedefaultmidpunct}
	{\mcitedefaultendpunct}{\mcitedefaultseppunct}\relax
	\EndOfBibitem
	\bibitem[Collins \latin{et~al.}(2013)Collins, Sambath, Harris, and
	Basaran]{collins13}
	Collins,~R.~T.; Sambath,~K.; Harris,~M.~T.; Basaran,~O.~A. \emph{Proceedings of
		the National Academy of Sciences} \textbf{2013}, \emph{110}, 4905--4910\relax
	\mciteBstWouldAddEndPuncttrue
	\mciteSetBstMidEndSepPunct{\mcitedefaultmidpunct}
	{\mcitedefaultendpunct}{\mcitedefaultseppunct}\relax
	\EndOfBibitem
\end{mcitethebibliography}
\end{document}